# UNDERSTANDING CUSTOMERS' EVALUATIONS THROUGH MINING AIRLINE REVIEWS

Ibrahim Yakut[1], Tugba Turkoglu[1] and Fikriye Yakut[2]

[1]Department of Computer Engineering, Anadolu University, Eskisehir, Turkey
[2]Department of Civil Aviation Management, Anadolu University, Eskisehir, Turkey

## ABSTRACT

*Data mining can be evaluated as a strategic tool to determine the customer profiles in order to learn customer expectations and requirements. Airline customers have different characteristics and if passenger reviews about their trip experiences are correctly analyzed, companies can increase customer satisfaction by improving provided services. In this study, we investigate customer review data for in-flight services of airline companies and draw customer models with respect to such data. In this sense, we apply two approaches as feature-based and clustering-based modelling. In feature-based modelling, customers are grouped into categories based on features such as cabin flown types, experienced airline companies. In clustering-based modelling, customers are first clustered via k-means clustering and then modeled. We apply multivariate regression analysis to model customer groups in both cases. During this, we try to understand how customers evaluate the given services and what dominant characteristics of in-flight services can be from the customer viewpoint.*

## 1. INTRODUCTION

Data mining covers a variety of techniques to discover inherent but meaningful patterns from huge amount of data. Using such patterns, some models can be constructed to facilitate business processes such as decision-making, investment planning, marketing strategy development and so on. Airline companies in current competitive business environment can improve suitable strategies to maintain the highest level of customer satisfaction and provide high quality service. To achieve such quality service, first of all, what *passengers*, 'customers' preferred for that term along this text regard to management sciences, expect from airline services are need to be analyzed and specified. In this study, we want to answer "what factors are dominantly essential while customers are rating the services provided by company?" We investigate ratings given by customers after their flight experiences. Over given a couple of ratings for each flight, we try to understand what factors are more important in the view of customers. We group the customers and draw models for each group. By the way, we demonstrate extensive empirical analysis results on real customer review data.

The paper is organized as follows: Section 2 provides a brief survey of the related studies in the area while In Section 3; we introduce data analysis and modelling methodology used in this work. In Section 4, after introducing airline customer review data, we exhibit our empirical findings using our methodology. Finally, we draw conclusions and give future research directions in the last section.

## 2. RELATED WORK

Data mining techniques can effectively be used by airlines for developing strategies about customers through marketing strategy. In this process, fundamentally these topics can be listed under four issues as customer value measurement, customer retention, customer growth and customer acquisition [1]. There are several studies concerning about such issues for promoting airline services with higher level of customer satisfaction. In this context, Liou and Tzeng [2] explore customer behavior in the Taiwan airline market in the context of questionnaire responses for single and multiple-choice answers. Then they evaluate customer response with rough set-based classification and they find that two criteria dominate customer decision-making: safety and price. Miranda and Henriques [3] point out that companies need to different customer segmentation framework in order to define different campaign strategy after analyzing the airline customer data by clustering. The authors evaluate three different clustering algorithms as of $k$-means, SOM and HSOM performance evaluated. According to analyze results based on data with demographic, flight preferences and history features belong to 20,000 customers. They concluded $k$-means is the best where SOM technique gives similar results. Our study can be considered similar to so mentioned ones in clustering customers based on personal data.

There are some studies investigating frequent flyer passengers through improvement for airline customer relation management. Maalouf and Mansour [1] use frequent flyer passenger data from over 1 million passenger activity records taken from about 80,000 passengers for 6 year periods. They evaluate such records using clustering and association rules to develop marketing and management strategies to improve customer relationships. Yan et al [4] use ID3 classification algorithm in order to analyze airport passenger survey data. They indicate that airlines desiring to develop customer process have to form marketing strategies considering frequent flyer passengers in both present and future. In this study, our concentration is to understand customer priorities about in-flight services. By analyzing customer priorities on their reviews, we model their attitudes using regression analysis to improve customer relationship management and develop marketing strategies.

## 3. METHODOLOGY: DATA ANALYSIS AND MODELLING

In this section, we introduce how to model passenger attitudes based on given review about airline in-flight services. In our study, customers are modelled in two different approaches as seen from Figure 1. Our first approach is based on dividing users having the same features in the same group. Feature can be one of airline travelled, cabin flown either economy or business and, so on. After selecting which feature is proper, then customer review data are grouped according to such feature values of customer. In our second approach, we apply clustering to group of users before drawing regression models. Since clustering especially $k$-means clustering which we are going to use is very effective with principal component analysis [5]. We apply such method before clustering customers.

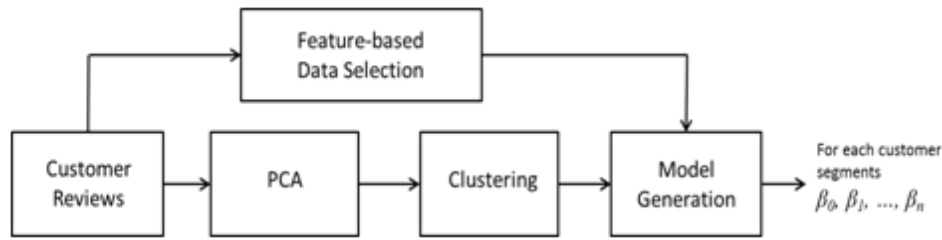

Figure 1.Model generation based on two different approaches

## 3.1. Principle Component Analysis

Principal component analysis (PCA) is one of the most popular multivariate statistical techniques and it is a well-known dimensionality reduction procedure [6]. Since focusing on to preserve correlation and variance between attributes[7], PCA is also effective for getting rid of noise effects on data. The superiority of PCA can be summarized as extracting the most important information from dataset, compressing the size of the dataset by keeping only this important information, simplifying the description of the dataset which enables low dimensional visualizations, analyzing the structure of the observations and the variables.

Six general steps for performing a principal component analysis can be listed as

    **Input:** Whole dataset $D$ consisting of $n$-dimensional samples
    **Output:** Reduced dataset $W$ with $p$ dimensions
        **Step 1:** Take the ignoring the class labels.
        **Step 2:** Compute the $n$-dimensional mean vector
        **Step 3:** Compute the scatter matrix of the whole dataset
        **Step 4:** Compute eigenvectors and corresponding eigenvalues
        **Step 5:** Sort the eigenvectors by decreasing eigenvalues and choose $p$ eigenvectors with the largest eigenvalues to form a $d$ x $p$ dimensional matrix $W$
        **Step 6:** Use $W$ matrix to transform the samples onto the new subspace.

The input matrix $D$ has $m$ rows (entities) and $n$ columns (features) where $n>p$. In our data model entities are customers and features are review values in $n$ dimension. After applying PCA, the output matrix is $D'=DW^T$ having $m$ rows and $p$ columns where superscript $T$ stands for transpose of a matrix. By the way, $n$ features are reduced to $p$ features which are effectively represent the original data features. We prefer to apply PCA before clustering analysis since not only recent results shows that PCA provides effective results with $k$-means but also but also visualize our findings on two-dimensional space [5, 8].

## 3.2 Clustering

Clustering is an unsupervised learning method decomposing a given set of objects into subgroups (i.e. clusters) based on object similarities[ 9]. The objective is to divide the data set in such a way that objects belonging to the same cluster are as similar as possible whereas objects belonging to different clusters are as dissimilar as possible[10]. $k$-means is one of the simplest and most commonly used unsupervised learning algorithm that solve the clustering problem in the first approach[10]. Algorithmic steps for $k$-means clustering for our case where entities are customers and $k$ reduced features of reviews can be shown below:

**Input:** $W$, $k$
**Output:** Cluster indices of $n$ customers
    **Step 1:** Randomly select $k$ cluster centers.
    **Step 2:** Calculate the distance between each customer and cluster centers using Euclidean distance.
    **Step 3:** Assign each customer to the cluster center so that there is minimum distance between the customer and the cluster center.
    **Step 4:** Recalculate new cluster centers.
    **Step 5:** Recalculate the distance between each customer and new obtained cluster centers.
    **Step 6:** If no data point is reassigned then stop, otherwise repeat Step 3-5.

### 3.3 Regression Analysis

Regression analysis is widely used for prediction and also understand the estimating the relationships among variables. Variables are divided into two as dependent and independent variables while there are single and multivariate regression analyses. If data is going to be analyzed using a single independent variable, it is called one variable regression; otherwise it is called multivariate regression analysis [10]. Multivariate regression analysis is divided into two as standard and hierarchical regression analysis. We used standard multivariate regression analysis because we had more than one independent variable for our study. Our goal is to perform multivariate regression analysis on relationships between independent and dependent variables. Multivariate regression analysis formula can be given below:

$$y = \varepsilon + \beta_0 + \sum_{j=1}^{n} \beta_j x_j. \qquad (1)$$

where $x_j$, $y$, $\beta_j$ and $\varepsilon$ are independent values, dependent values, parameters. The estimated $y$ value is calculated by the values obtained from the regression results and error term ($\varepsilon$) obtained by finding absolute values of difference of the actual $y$ values from the estimated $y$ values. $n$ value is number of independent values.

## 4. MODELLING PASSENGERS ACCORDING TO IN-FLIGHT REVIEWS

In this section, first of all we introduce customer review dataset then model the customers with respect to given methodology in the previous section.

### 4.1 Customer Review Data

Skytrax, a consultancy firm located in London, United Kingdom, conducts research and consultancy mostly within the aviation sector [11]. This company conducts research for airlines to find the best cabin staff, airport, airline, airline lounge, in-flight entertainment, on board-catering and several other elements of air travel. Based on conducted survey and their evaluation methodology, they give world airline awards and announce annual airline rankings every year. In this study, we use Skytrax's airline rankings which are publicly available in Skytrax's interactive web site, www.airlinequality.com. There are latest travel reviews and customer trip ratings use for 681 airlines and 728 airports across the world. In our study, we use customer review data for selected airlines. Example of customer review data is demonstrated in Figure 2 and there are comments about experienced travel and numerical ratings represented either with stars from 1 to 5 or bar from 1 to 10.

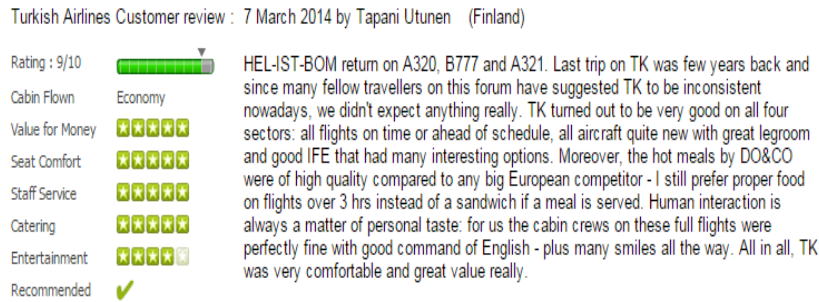

Figure 2. Customer reviews about the in-flight services.

Our study concentrates such numerical ratings with other data given in left-side of Figure 2 rather than comments in the right. Such ratings are also useful information that customer has been flown in economy, premium economy or business class and whether customer recommend or not recommend the airline company to potential customer audience. Considering huge amount of reviews in Skytrax, we want to reduce input data selecting a convenient subset consisting of Star Alliance largest members. To minimize the variations due to different airlines, we prefer to use data related to airlines from the same alliance. Airline alliance is an agreement between two or more airlines to cooperate on a substantial level. When an airline joins an alliance, the reliability of its services offered to customers not only depends on the flights the airline operates, but also on the operations of the rest of the alliance members [12]. Airlines which are members of the same alliance, they usually handle same standards and they have similar characteristics. Choosing the airlines from same alliance is important for this reason. Star Alliance is the global airline network with the highest number of airlines, daily flights, destinations and countries flown to [13, 14]. In this study, input data is costumer review data given from 1 January 2014 to 31 December 2014 and the reviews are about 5 airline companies that member of star alliance. These airlines are United, Lufthansa, Air China, Turkish, All Nippon Airways (ANA). Chosen airlines are the star alliance's largest members according to number of passengers carried per year [13]. These airlines are from different regions such as North America (*United* from USA), Europe (*Lufthansa* from Germany), Middle East (*Turkish* from Turkey), and Far East (*Air China* from China and *ANA* from Japan). Customers from different regions of world may reflect worldwide and regional characteristics of customer behavior. Now, we are going to model customers with the hypothesis that *rating* value given by customer is dependent to sub-rating values such as *value for money*, *seat comfort*, *staff service*, *catering* and *entertainment*. Based on that hypothesis, there are a dependent variable (*y*) as a rating and 5 independent variables ($x_j$s) as sub-ratings according to (1). Consequently in our models, $\beta_j$ values represent value for money (*j*=1), seat comfort (*j*=2), staff service (*j*=3), catering (*j*=4), and entertainment (*j*=5), respectively. As a minor issue to regulate the ranges, we doubled sub-ratings in from 5 unit interval to 10 unit interval as ratings ranging from 1 to 10.

## 4.2 Feature-based Modelling

In feature-based modeling, we group customers according to features such as cabin flown and airlines traveled. In this subsection, we demonstrate obtained models from such feature-based

International Journal of Data Mining & Knowledge Management Process (IJDKP) Vol.5, No.6, November 2015

data grouping. In the first case, we group customer data into three groups as business, economy and premium economy in which there are 381, 1002 and 111 customers, respectively. After performing regression analysis for each group, we obtain $β_j$ values as given in Table 1. Then we plot such values in Figure 3 and upcoming figures discarding $β_0$ values since our concentration is the relation with ratings and sub-ratings. According to Figure 3, we observed that for all flown class value of money is the most important factor. Looking all flown class together, staff service can be considered second important factor after value of money. Obviously, seat comfort is not essential factor for premium economy class, while it has nearly the same effect business and economy class customers. About Figure 3, we can also say that business class customer's catering expectation is higher than others but entertainment expectation is lower than other classes. Figure 3 also shows that premium economy class customers give entertainment more importance than other customers uses other cabin flown and even it is prior to seat comfort in contrast to other customers. $β_0$ values which we found regression operation did not show on the graph because $β_0$ values are too small enough to affect the result.

Table 1. The Distribution of Beta Values to Each Cluster

|  | $β_0$ | $β_1$ | $β_2$ | $β_3$ | $β_4$ | $β_5$ |
|---|---|---|---|---|---|---|
| Sub-rating | - | Value of Money | Seat Comfort | Staff Service | Catering | Entertain. |
| Business | -0.2349 | 0.9803 | 0.1227 | 0.3696 | 0.3895 | -0.0119 |
| Economy | -0.3042 | 0.8985 | 0.1082 | 0.5839 | 0.0725 | 0.0960 |
| Prem. Eco. | -0.2762 | 1.0415 | -0.4260 | 0.6712 | 0.2390 | 0.2666 |

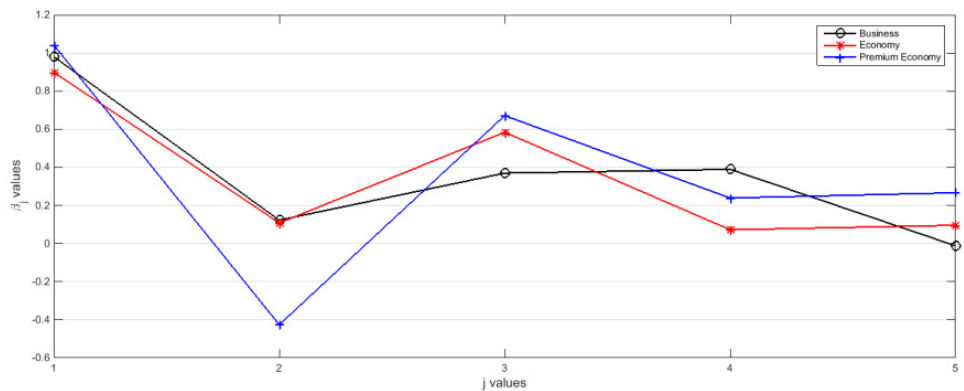

Figure 3. Relation between cabin flown and model parameters.

Our second case is to group users with respect to airline traveled. In sampled review data, there are 545 United, 434 Lufthansa, 98 Air China, 341 Turkish, and 76 ANA passengers and our groups consisting of such passengers. After applying regression analysis, we give our models in Figure 4. According to such figure, again value of money is the considerably essential factor according to all groups. Value of money has the most effect for United passengers in comparison to other airline passengers and but for them seat comfort has the most little effects both in their overall and in contrast to other airline passengers. According to the Figure 4, we can say that Lufthansa passengers' sub-rating weights are all closer except entertainment. The outstanding feature of Lufthansa group, overall rating is based on sub-ratings in significantly balanced way. Interestingly from Figure 4, passengers who traveled with Air China and Turkish Airlines have similar tendencies about their reviews except entertainment factor. Entertainment is more important for Air China passengers than those of Turkish (THY). Looking at Figure 4, we can remark that for ANA passengers the most important factor is seat comfort even slightly better than value of money. Unlike other airline companies, for ANA staff service has fewer effects on overall ratings. Thus, staff service importance are lower for Japanese which may cause from that the Japanese typically demand much higher levels and quality of service than expected elsewhere in the worldwide as stated in [14] and service is default must for Japanese passengers. From Figure 4, we can also note that for ANA passengers, catering has also significant effect in overall.

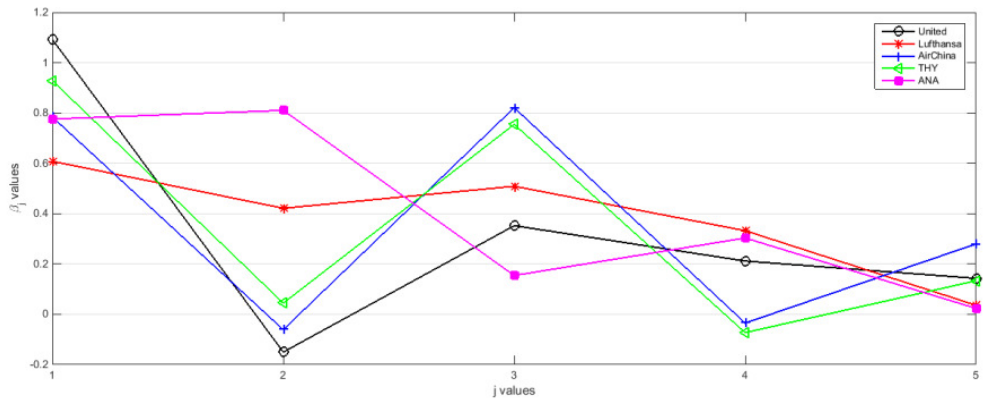

Figure 4. Relationship between airlines traveled and model parameters.

## 4.3 Clustering-based Modelling

As we mentioned in Section 3, initially we reduced our data with six-parameters (5 sub-ratings + 1 rating) into 2-dimensional by using PCA. Then we applied *k*-means with regression analysis to identify how many cluster would be appropriate for us. For this reason, we select *k* values from 1 to 10 to obtain regression errors ($\varepsilon$) and repeated this process 100 times. Note that along those trials, we evaluate such errors in context of mean absolute errors as used in [8]. We display obtained average $\varepsilon$ values in *y*-axis corresponding *k* values in *x*-axis in Figure 5. The figure shows that there are local minimum at *k*=2 and it seems errors get decrease with increasing values after *k*=6. We prefer to use *k*=6, since we can consider there is acceptable regression error and 6 clusters are feasible to visualize and discuss effectively. There is less regression with *k*=2 comparing to *k*=6 but less representative clusters and for higher *k* values, effective analysis, modelling and discussion may be complex. Hence, for the sake of simplicity and acceptable modelling accuracy, we prefer to 6 clusters to model customers.

After determining proper cluster number, there is still one issue to fix up to achieve stable results and repeatability of results. While initial cluster centers can be randomly determined, they can be selected proper values. We consider latter case and determine six center points; the four of them are the rectangular corners consisting of combinations of maximum and minimum values of each column of reduced data while the remaining two points are quarter distance points when the corners grouped in pairs. As a result center points which we have already selected are {(7, 9), (- 14, 9), (7, -7), (-3.5, -3), (-3.5, 5), (-14, -7)}. Then, after setting up initial parameter, we execute *k*-means algorithm then obtain results as pictured in Figure 6. According to this figure, there are distributed points each representing a customer within six clusters with final center points shown with cross sign (×).

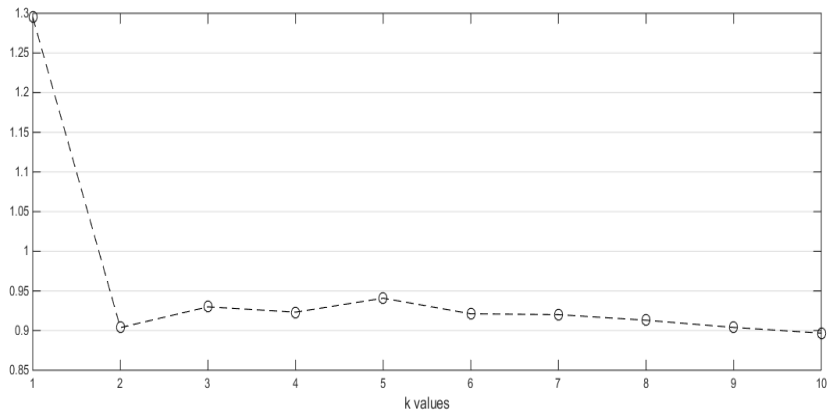

Figure 5. Regression errors with respect to varying cluster numbers

Based on clusters in depicted in Figure 6, we plot regression parameters belong to clusters in Figure 7. We also find and give useful details about customers in each cluster in Table 2. Note that we prefer to label recommended user as satisfied users as in last column in Table 2 since customer satisfaction is the determinant of intention to recommend [15].According to Table 2, first of all about Cluster1, we can say that percentage of business class passengers is have the largest in this cluster than the other clusters while it is the most crowded cluster with 25.4% of all customers investigated. This cluster has the highest overall rating value with 6.58 and there are about 70% of satisfied customers. From Figure 7, Cluster 1 has closer weighted sub-ratings effect on the overall effect which is similar to Lufthansa passengers. According to Figure 7, we can say about Cluster 2 passengers similar attitudes towards ratings with Cluster 5. Considering average overall ratings in Table 2 with the figure, difference between two graphs is shift due to average overall ratings rather than model-oriented difference. There are 25.3% of total customer by summing customers in Cluster 2 and 5; however business customer rate of both is half of Customer 1. Customers in this combined cluster have priorities ordered value of money, staff service and entertainment and the other factors are not essential for such kind of customers. According to Figure 7, Cluster 3 costumers give importance to staff service and catering at least as much as value of money. For Cluster 3, the staff service makes sense but they do not care about entertainment when evaluating overall performance of given service. According to Table 2, the percentage of Cluster 3 population is 13.7% of overall. One outstanding cluster is number 4 where there are customers caring about entertainment as much as value of money and the other factors are not so vital for them during in-flight services. However, looking at Table 2, only 53% of satisfied users of that cluster and they are 11.8% of all customers. Finally, as shown in Table 2, Cluster 6 is the one of most crowded clusters which takes individually about 23.8% of all customers. Its tendencies are similar with unsatisfied customers depicted in Figure 5 which can be supported with just 34.8% of recommended customers in Cluster 6 given in Table 2, as well.

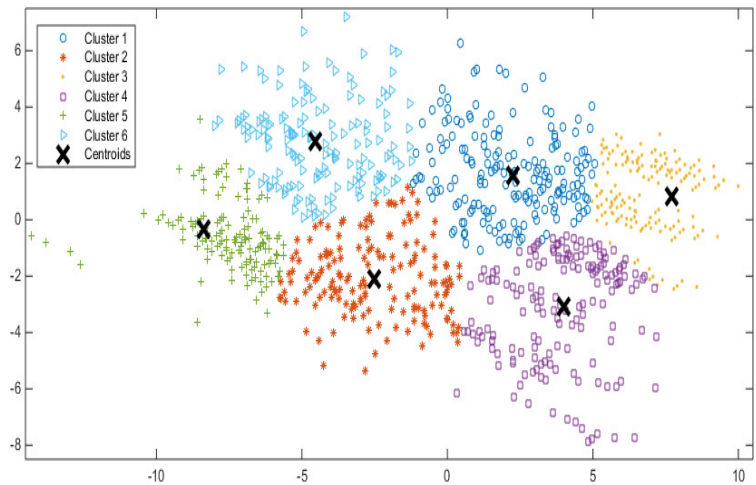

Figure 6. The distribution of the clusters.

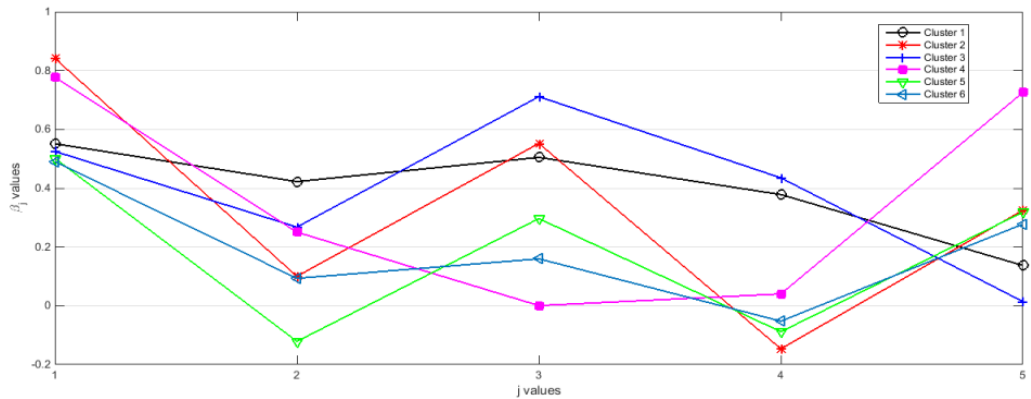

Figure 7. Model parameters of each cluster

Table 2. Cluster Statistics

| Cluster # | Total No. Pasngr | No. of Busns | No. of Econ. | No. of Prem. Eco. | Aver. Rating | Satsf'd Cust. (%) |
|---|---|---|---|---|---|---|
| Clust. 1 | 379 | 149 | 213 | 17 | 6.58 | 69.7 |
| Clust. 2 | 206 | 41 | 150 | 15 | 5.85 | 62.6 |
| Clust. 3 | 204 | 55 | 135 | 14 | 6.29 | 71.1 |
| Clust. 4 | 177 | 50 | 111 | 16 | 5.10 | 53,1 |
| Clust. 5 | 172 | 29 | 132 | 11 | 4.94 | 48.8 |
| Clust. 6 | 356 | 57 | 261 | 38 | 3.86 | 34.8 |

## 5. CONCLUDING REMARKS AND FUTURE WORKS

It is critical for companies to collect customer data revealing hints about their attitudes about their sector and process such data in order to extract useful patterns worth to be considered during strategy development especially for marketing and customer relations management. In this study, we show how to mine available Skytrax customer review data about in-flight services in the sense of customer perceptions. We show that some inferences can be captured to understand how customers are evaluating the given services when properly modelling customers. Based on our analysis, it can be stated that customers differently perceive sub-rating values and return overall ratings based on such perceptions. According to some customer groups, value of money is very essential and for some others, staff service is the dominant factor for service evaluation.

This study offers data analysis and modelling framework for promoting airline in-flight services with the discussion of current outcomes. There are still issues to be examined in future studies. For example, initially we select smaller sample of reviews. However, different data selection can be realized in order to reach outcomes reflecting some other features. Another issue is that there are many data mining and regression techniques in addition to k-means clustering and multivariate regression analysis. Exploiting other clustering algorithm may present different grouping of customers worth to be analyzed. We believe that there will be fruitful discussions through different kind of data mining applications for this data.

## ACKNOWLEDGEMENTS

This study was supported by Anadolu University Scientific Research Project Comission under the grant no: 1502F062.

**AUTHORS**

Ibrahim Yakut is an Assistant Professor in Dept. of Computer Engineering in Anadolu University, Eskisehir, Turkey. He received his M.S. and PhD. degrees in Computer Engineering from Anadolu University in 2008 and 2012, respectively. He was a one-year visiting research fellow at MSIS Dept. of Rutgers University, Newark, USA in years 2013-2014. His research interests are recommender systems and privacy-preserving data mining.

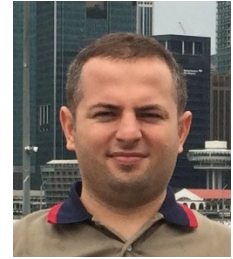

Tugba Turkoglu is a Research Assistant in Department of Computer Engineering, Anadolu University,, Turkey. She received the B.S. degree in Computer Engineering from Suleyman Demirel University, Isparta, Turkey in 2013 and is still master's candidate in Computer Engineering in Anadolu University, Eskisehir, Turkey. Her area of research is centered around data mining and programming.

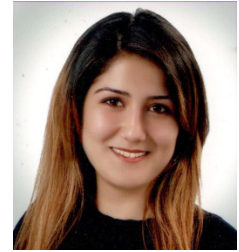

Fikriye Yakut is a Lecturer in Department of Transportational Services in the School of Advanced Vocational Studies, Bilgi University, Istanbul, Turkey. She received both B.S. and M.S. degree from Department of Civil Aviation Management, Anadolu University, Eskisehir, Turkey while being also PhD. Candidate at at the same department . Her areas of research include marketing and management in civil   aviation business.

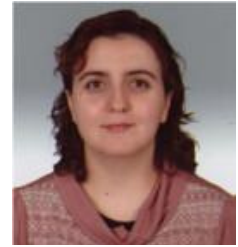